\documentstyle[aps,preprint]{revtex}
\begin{document}
\draft
\title{Self-Organized Criticality in Non-conserved Systems }
\author{A. Alan Middleton and Chao Tang}
\address
{NEC Research Institute, 4 Independence Way, Princeton, NJ 08540}

\date{\today}
\maketitle

\begin{abstract}

The origin of self-organized criticality in a model without conservation
law (Olami, Feder, and Christensen, Phys. Rev. Lett. {\bf 68}, 1244
(1992)) is studied.  The homogeneous system with periodic boundary
condition is found to be periodic and neutrally stable.
A change to open boundaries results in the invasion of
the interior by a ``self-organized'' region.  The mechanism for the
self-organization is closely related to the synchronization or
phase-locking of the individual elements with each other.
A simplified model of marginal oscillator locking on a directed
lattice is used to explain many of the features in the non-conserved model:
in particular, the dependence of the avalanche-distribution exponent
on the conservation parameter $\alpha$ is examined.

\end{abstract}

\newpage
\narrowtext

The phenomenon of the self-organized criticality (SOC) \cite{btw} is
characterized by spontaneous and dynamical generation of scale-invariance
in an extended non-equilibrium system.  One of the key issues in this
field is to identify the
mechanism(s) of SOC.  It has been conjectured that conservation laws or
special symmetries are necessary \cite{grinstein}.
Conservation laws certainly are of great importance in ``sandpile''
models \cite{btw,kadanoff,ccgs}, where the scale invariance can be shown
to follow from a local conservation law (sand grains are conserved except
at the boundaries of the pile) \cite{dharetal}.  In this sense, the
origins of long range correlations in SOC systems with conservation are
well understood, though not all exponents have been calculated
analytically.  However, models have been constructed
\cite{life,feder,fire,ofc} that have no apparent conservation law, and
yet display a power-law distribution of avalanche sizes.  Of particular
interest is the model proposed by Olami, Feder, and Christensen (OFC)
\cite{ofc,socolar,grassberger} which, they argue, models earthquake
dynamics.  The OFC model is very similar to the conserved sandpile model
\cite{btw}, but it has a parameter which defines the degree of
conservation.  In this paper, we study in detail the OFC model and we
find that the self-organization is due to synchronization or
``phase-locking'' -- a mechanism very different from that in
the conserved models.

In the OFC model, dynamical ``height'' variables $h_{i}$ are defined
on sites $i$ of a square lattice.
The $h_{i}$ increase at unit rate until $h=1$
at some location.  The site $j$ where $h_{j}=1$ is considered to be
unstable and will ``topple''.  The rule of toppling is that
when $h_{j} \geq 1$, then $h_{j} \rightarrow 0$
and $h_{k} \rightarrow h_{k} + \alpha h_{j}$, for all $k$ neighboring $j$.
The toppling on site $j$ may cause its neighbors to become unstable
($h_{k} \geq 1$) and to topple.  This procedure is repeated until all
sites are stable ($h_{i} < 1$ everywhere).   The magnitude of
the avalanche is given by the total ``energy'' dissipated in the
process, i.e., the total change in $\sum_{i} h_{i}$.  The avalanche,
which happens instantly on the time scale of driving \cite{septime}, is
then followed by growth.  The parameter $\alpha$ is
the measure of conservation (of $h$'s).  When $\alpha = 1/4$,
the model is conserved and it is in the same universality class
as for the BTW model \cite{btw,ofc,univ}.  We study here the
non-conservative case $0<\alpha<1/4$.

We note that there is an
ambiguity in this model.  After some number of avalanches, it will
occur that more than one site will have exactly the same height,
as more than one site may topple to a height of exactly zero in a
single avalanche.  This can lead to two neighboring sites toppling
simultaneously; the above procedure does not define the result of such
events.  We have modified the rules in several ways, e.g. by adding a
very small amount of random noise to ensure that no two sites have the
same height, and they all give the similar results.

As most avalanches are small, and do not change the $h_{i}$ at many
sites, we do not need to scan the whole lattice after each avalanche
in order to determine the most unstable site.  We use a tree structure
to keep track of the highest values of $h_{i}$ in order to determine
avalanche trigger sites.  Using this technique, we have simulated up
to more than $10^{10}$ avalanches for single systems.

We find that the system with periodic boundary conditions quickly
reaches an exactly periodic state \cite{socolar}, with a unique
period in the slow time.  It has been noted \cite{ofc,grassberger}
that in the case of periodic
boundary conditions, the avalanche size distribution function drops
very quickly with size.  In the case of our modified model, which
prevents two sites from having the same value of $h$, all avalanches
consist of the toppling of exactly one lattice site, after a brief
transient time.  In a periodic state, the $h_i$'s take turns
to topple, one by one.  The height $h$ decreases
by one each time a site topples and increases by $4\alpha$
due to the toppling of the four neighbors when they reach $h=1$.
Thus the period of all these periodic states is $1-4\alpha$ in
the slow time variable, so that the slow
``growth'' is balanced exactly by the dissipation due to toppling.
These periodic states are highly degenerate and neutrally stable
in the sense that a typical small perturbation of the height at a single
site in
a periodic state is still a periodic state.  They are similar to
the neutrally stable periodic states in coupled
oscillators \cite{wiesenfeld}.  There is a continuous
set of periodic states in the attractor, with measure
$(1-4\alpha)^{V}$ in the initial phase space, where $V$ is the
system volume.

Any inhomogeneity, such as a change in boundary conditions,
destroys such simple periodic states.
When the boundary conditions are open, the system can no longer have
period $1-4\alpha$, as the boundary sites have 3 neighbors (we
study a system that is open on one axis, with the other
directions periodic).  Initially, the interior sites quickly
converge to a nearly periodic state and topple with
period $1-4\alpha$, but the boundaries are aperiodic.  At longer
times, the aperiodic region invades the periodic interior, as shown in
Fig.\ \ref{invasionfig}.  This invasion, which destroys the
periodicity in the interior and builds up long-range correlations,
occurs by a mechanism similar to oscillator locking,
as we describe below.  The interface
between the two regions is well defined on scales larger than one
or two lattice constants.  The invasion distance appears to have
a power law dependence on time, $y(t) \sim t^{\beta(\alpha)}$
as shown in Fig.\ \ref{invadepowerfig},
with $\beta = 0.23\pm 0.08, 0.63\pm0.08$ for $\alpha = 0.07, 0.15$,
respectively.  We see such invasion occurring even for values of
$\alpha < 0.05$, though $\beta$ appears to be quite small, so that the
invasion is extremely slow.  This suggests that the transition to
non-SOC behavior claimed by Olami, Feder, and Christensen \cite{ofc}
may only be apparent, due to the finite time of the simulations; we
note that the time for complete invasion of a $128^{2}$ system with
$\alpha = 0.07$ is greater than $10^{10}$ avalanches.  In the limit
of long times, when the invasion crosses the whole sample, the
distribution $P(s;\alpha)$ of avalanches of size $s$ is
a power-law, with
\begin{equation}
P(s;\alpha) \sim s^{-\tau(\alpha)}.
\end{equation}
Consistent with OFC, we find $\tau(\alpha) = 3.2 \pm 0.1,
2.3\pm 0.1$ for $\alpha = 0.07, 0.15$, respectively.
The avalanches are not uniformly distributed in the system;
the typical avalanche size grows with distance from the edge.

The influence of inhomogeneity, which leads to a breaking of the
$1-4\alpha$ periodic state, is quite strong. Even a single defect,
where $h$ is set to zero for all time, destroys the periodic state,
and leads to a power law distribution of avalanche sizes.

The spatial distribution of length scales is apparent in
Fig.\ \ref{invasionfig}(d),
which shows an example of a configuration between avalanches in the
steady state.  Near the boundaries, the $h_{i}$ have only short range
correlations, but this correlation length grows with distance from the
boundary.  The {\em typical} avalanches in the interior, though, still
involve the toppling of only one lattice site.  Infrequently, an
avalanche is triggered near the boundaries, which penetrates into a
longer-range correlated region, giving a large avalanche.  We define a
toppling rate $r(y)$ as a function of distance $y$ from the boundary,
which gives the inverse of the mean time between topplings.  At the
boundaries, where the sites have only three neighbors, $r(y)$ is
smaller than the interior, where $r(y) \rightarrow (1-4\alpha)^{-1}$
as $y\rightarrow\infty$.  The toppling rate differential, defined as
$\delta r(y) = (1-4\alpha)^{-1} - r(y)$ is found to behave as
\begin{equation}
\delta r(y) \sim y^{-\eta}
\label{drate}
\end{equation}
with $\eta = 3.2\pm0.6, 1.8\pm0.2$ for $\alpha = 0.07, 0.15$,
respectively.  Let $h_t(y)$ be the average height just before toppling,
$R(y) \equiv (1-4\alpha)r(y)h_t(y)$ is then the dissipation rate and
$R(y) \rightarrow 1$ as $y\rightarrow\infty$.  It can be shown that
in the steady state $\delta R(y) = 1 - R(y)$
behaves as $\delta R(y) = exp (-y\sqrt{\frac{1-4\alpha}{\alpha}})$.  Thus
the dissipation rate is rather uniform except within a boundary layer of
thickness $\sqrt{\frac{\alpha}{1-4\alpha}}$.  The power-law behavior of
Eq.~(\ref{drate}) must be compensated by a power-law in $\delta h_t(y) =
h_t(y)-1$:
\begin{equation}
\delta h_t(y) \sim y^{-\eta}.
\label{dht}
\end{equation}
Note that the toppling height can be larger than 1 only when the toppling
is triggered by neighboring sites and hence is part of an avalanche which
involves more than one site.  The picture we get from Eqs.~(\ref{drate})
and (\ref{dht}) is that although the toppling happens less frequently as
we move towards the boundary a larger portion of it is due to
multi-site avalanches.

In order to gain some insights on the build-up of long range correlations
in the inhomogeneous system, we consider a system which has only two sites:
$h_1$ and $h_2$.  Let us first consider the homogeneous case: both
$h_1$ and $h_2$ are driven with unit rate and when one of them reaches
the value one it topples.  The rule of toppling is that: if $h_{1(2)}
\ge 1$, then $h_{2(1)} \rightarrow h_{2(1)} + \alpha h_{1(2)}$ and $h_{1(2)}
\rightarrow 0$.  This simple system is completely integrable.  It has a
continuous set of periodic states which are marginally stable.  To
illustrate its dynamics, we construct a Poincar\'{e} map.  Denote
$h_1(n)$ to be the value of $h_1$ right after the $n$th toppling of
$h_2$.  It is easy to show that \cite{socolar}
\begin{equation}
h_1(n+1) = \left\{ \begin{array}{ll}
             h_1(n)                       & \alpha \le  h_1(n) < 1 \\
             1+\alpha-\alpha h_1(n) \quad & 1 \le h_1(n) < 1/\alpha \\
             \alpha^2 h_1(n)              & h_1(n) \ge 1/\alpha
                   \end{array}
             \right.
\label{map-homo}
\end{equation}
which is sketched in Fig.\ \ref{mapfig}(a).  We see that there is a line
of marginally stable fixed points $h_1^* \in [\alpha,1)$.  These fixed
points are periodic states with period $1-\alpha$: $h_1$ and $h_2$ take
turns to topple and the toppling of one site will not trigger the
toppling of another ($h_1^* < 1$).  Now, we introduce a little inhomogeneity.
We drive $h_1$ with rate 1, but $h_2$ with a slightly slower rate
$(1+\epsilon)^{-1}$.  The Poincar\'{e} map now reads
\begin{equation}
h_1(n+1) = \left\{ \begin{array}{ll}
  h_1(n)+\epsilon (1-\alpha)                      & \alpha \le  h_1(n) < 1 \\
  1+\alpha+\epsilon-\alpha(1+\epsilon)h_1(n)\quad & 1 \le h_1(n) < 1/\alpha \\
  \alpha^2 h_1(n)                                 & h_1(n) \ge 1/\alpha
                   \end{array}
             \right.
\label{map-inhomo}
\end{equation}
which is sketched in Fig.\ \ref{mapfig}(b).  In this case, there is
only one fixed point $h_1^* = 1+\frac{1-\alpha}{1+\alpha}\epsilon$.
This fixed point is the phase-locked or synchronized state: the
toppling of $h_2$ will trigger $h_1$ to topple ($h_1^* > 1$).  Note that
Eq.~(\ref{map-inhomo}) is only $\epsilon$ away from
Eq.~(\ref{map-homo}), so the locking is rather weak and fluctuations can
play a crucial role.  In the OFC model with open boundary conditions,
the boundaries introduce inhomogeneity.  The sites at the open
boundaries have only three neighbors
and hence have a {\em slower effective
growth rate}.  This inhomogeneity in the effective growth rate propagates
into the interior of the sample, causing phase-locking and thus long
range correlation.  However, the whole system is not in a synchronized
state.  Rather, it is only ``marginally'' locked so that it gives a
power-law distribution of avalanche sizes.

We do not have a complete theory for the emergence of the ``marginal
locking'' in the OFC model. However, we can
abstract some of the features to
construct a simpler model which also exhibits SOC.  This model is
defined on a directed lattice to simplify avalanches.  We define
dynamical variables $0 \leq \phi_{i} < 1$ as the {\em phase} of the
next toppling time of
a site;
this phase is related to the height
$h_i$ in the OFC model at a fixed time, modulo the natural period
($1-4\alpha$).
At random sites on the boundary, we initiate
a toppling which
changes the phase according to
$\phi_i \rightarrow \phi'_i = \phi_i + \alpha$.
If the
phase $\phi_j$ at a neighboring ``downhill''
site $j=i+\hat{x}$ or $i+\hat{y}$ (see
inset in Fig.\ \ref{simplesnap}) meets the locking condition
$(\phi_i - \phi_j \bmod 1) < \alpha$,
the site $j$ locks onto the boundary site, with $\phi_j
\rightarrow \phi'_i$.  This disturbance can then continue to propagate by sites
further from the boundary becoming locked, in zero time.
We refer to this locking
as ``marginal'' since neighboring phases only lock upon a {\it crossing} in
the toppling time; the configuration is neutrally stable with respect to a
continuous set of perturbations.
This model differs from the OFC model most notably in the directed lattice
and the instantaneous locking (even though avalanches in OFC occur in zero
time, locking takes place over a time of order 1).

A snapshot of the $\phi_{i}$ in the steady state is depicted in
Fig.\ \ref{simplesnap}(a).  The domains, bounded by solid lines,
are regions where the toppling times are identical.
Fig.\ \ref{simplesnap}(b) shows a configuration after an avalanche,
with the dotted lines showing the previous domain configuration.
An avalanche crosses domain boundaries only when the
neighboring domains have times that differ by no more than $\alpha$.
Numerically, it is found that the number of domains $n(S)$ of size $S$
at a fixed time
has a power law tail that is independent of alpha, $n(S) \sim
S^{-\sigma}$, with $\sigma = 1.495\pm0.005$.  Yet the avalanche
distribution has an $\alpha$ dependent exponent, with the probability
of an avalanche of size $s$ behaving as $P(s)\sim s^{-\tau(\alpha)}$
(Fig.\ \ref{simpletau}).  Note that domains and avalanches are closely
related, with avalanches defining domains. The exponent $\sigma$ must
satisfy the bound $\sigma \leq 3/2$; the fact that our numerical result
saturates this inequality, to within numerical error, suggests that the
domains have linear roughness, with the average width of a domain proportional
to the length of the domain perpendicular to the boundary.

The relationship between $\tau$ and $\sigma$ can be approximately
explained.
Assuming that the toppling times of neighboring domains are independent
variables, the probability of an avalanche, which takes place in one
domain, incorporating a given neighboring domain is just $\alpha$.
The incorporation of a neighboring domain increases the avalanche size to
the scale of the neighboring domain, which is typically
larger than the original domain.
Given a distribution of domain sizes $n(S)\sim S^{-\sigma}$
and a scale independent probability $\alpha' \propto \alpha$
that a larger domain than the initial one
chosen will become part of the avalanche, followed by a probability
$\alpha'$ that an even larger domain will become part of the avalanche,
etc., leads to a distribution of avalanches $P(s)\sim s^{-\tau(\alpha)}$,
with $\tau(\alpha) = \sigma + \alpha'(1-\sigma)$.
This is in agreement with the linear fit of Fig.\ \ref{simpletau},
with $\alpha' = (1.20 \pm 0.04) \alpha$; the fitted value of
$\tau(0) = \sigma = 1.505 \pm 0.005$
agrees with the value determined by the domain distribution.
The varying exponent of the avalanche distribution is then explained by
the probability $\alpha'$ for the avalanche size
to move out further into the tails of the
$\alpha$-independent domain size distribution.

In this paper we have examined how SOC can arise in a model without a
conservation law.  The OFC model with periodic boundary condition
has a continuous set of neutrally stable periodic states.  In general,
inhomogeneity destroys these periodic states and causes phase-locking
which is the building block for long range correlations.  We found that
an open boundary results in the invasion of the interior by a marginally
locked region in which the avalanche size distribution is a power-law.
A simplified model on a directed lattice has been used to demonstrate
how an $\alpha$-dependent avalanche size distribution exponent can arise
in such non-conserved dynamical models.  Finally, we note that the OFC
model is similar to the coupled ``integrate-and-fire'' oscillators
studied in the context of neural networks and biology.  A close cousin
is the Peskin's model for the cardiac pacemaker. \cite{peskin}  We found
that the model in 2d with nearest neighbor coupling tends to lock into
some periodic or ``quasiperiodic'' cluster state and that it has
richer behaviors than simple synchronization.

\begin{figure}
\caption{
Configurations of the Olami-Feder-Christensen model at various
times after random initialization, demonstrating the invasion
of the short-range correlated interior by the ``SOC region''.
The density of the color corresponds
to the height variables $0 \le h_i < 1$.  The boundaries are periodic
in the vertical direction and open in the horizontal. The lattice
consists of $64^2$ sites, with $\alpha = 0.07$. Times are $t=$
(a) $1.2 \times 10^3$,
(b) $2.4 \times 10^3$,
(c) $6.0 \times 10^3$,
(d) $36.0 \times 10^3$.
}
\label{invasionfig}
\end{figure}

\begin{figure}
\caption{
A plot of the invasion distance vs.\ time for conservation parameters
$\alpha = 0.07, 0.15$. Power law fits are shown as dashed lines; symbols
indicate system size ($\triangle = 64^2$, $\Box = 128^2$,
$\bullet = 256^2$).
}
\label{invadepowerfig}
\end{figure}

\begin{figure}
\caption{
Return map for the two-site system: (a) homogeneous system; (b)
inhomogeneous system.
}
\label{mapfig}
\end{figure}

\begin{figure}
\caption{
Domains in the simplified SOC model described in the text, for a
$200^2$ directed lattice. (a) Configuration with domains of identical
toppling times indicated by solid lines. (b) Configuration after an
avalanche, with previous domains indicated by dashed lines. The inset
sketches the directed lattice.
}
\label{simplesnap}
\end{figure}

\begin{figure}
\caption{
The avalanche distribution exponent $\tau (\alpha)$ for the
directed model as a function of the conservation parameter $\alpha$.
The linear fit is discussed in the text.
}
\label{simpletau}
\end{figure}


\begin{references}
\bibitem{septime}
Note that there is an explicit separation of time scales in the model.
The slow time scale is the driving at unit rate.  The avalanches occur
instantaneously on this slow time scale.
\end{references}
\end{document}